# Multilayer plasmonic photonic structures embedding photochromic molecules or optical gain molecules

Francesco Scotognella[1,2]
[1] *Dipartimento di Fisica, Politecnico di Milano, Piazza Leonardo da Vinci 32, 20133 Milano, Italy*
[2] *Center for Nano Science and Technology@PoliMi, Istituto Italiano di Tecnologia, Via Giovanni Pascoli, 70/3, 20133, Milan, Italy*
*Corresponding author: francesco.scotognella@polimi.it

**Abstract**
We design photonic structures embedding different functional molecular systems of photochromic switching and lasing. We study the light absorption of two photochromic molecules and of 4,4'-bis[(N-carbazole)styryl]biphenyl (BSB-Cz) with density functional theory. For the photochromic diarylethene we derive the refractive index with Kramers-Kronig relations and we design multilayer photonic structures alternating diarylethene with either poly vinyl carbazole or fluorine indium co-doped cadmium oxide (FICO) nanoparticle-based layers. For BSB-Cz we design a one-dimensional photonic crystal alternating BSB-Cz layers wit $SiO_2$ layers and a microcavity embedding BSB-Cz between indium tin oxide (ITO) / FICO nanoparticle-based multilayers.

**Introduction**
The periodicity in one dimension of the refractive index, at a length scale comparable with the wavelength of light, can be easily obtained by stacking layers of two different materials, resulting in the so called Bragg mirrors, distributed Bragg reflectors, or Bragg stacks [1–3]. This kind of multilayers can be fabricated with different techniques, such as sputtering [4,5], plasma enhanced chemical vapor deposition [6] and spin coating [7]. A large variety of materials can be included as layers in one-dimensional multilayer photonic crystals [8]. For this reason, multilayer photonic crystals are promising for several applications [9,10]. The inclusion of plasmonic materials can enable wide electrochromic tuning [11]. Also the employment of photochromic molecule-based layers can be promising for colour switching devices, while the use of optical gain molecule-based lasers is a well-establish concept to fabricate distributed feedback lasers [12–14].

In this work we design multilayer photonic crystals and microcavities including two photochromic molecules reported in Refs. [15,16] and 4,4'-bis[(N-carbazole)styryl]biphenyl (BSB-Cz) [17] one of the most promising optical gain molecules. We study the optical properties of the three molecules with density functional theory. To extract the complex refractive index of diarylethene we use Kramers-Kronig relations. The multilayer photonic crystals are designed by alternating diarylethene with either poly vinyl carbazole or fluorine indium co-doped cadmium oxide (FICO) nanoparticle-based layers. Moreover, we engineer a one-dimensional BSB-Cz/$SiO_2$ photonic crystal and a microcavity embedding BSB-Cz between indium tin oxide (ITO) / FICO nanoparticle-based multilayers.

**Methods**
*Density Functional Theory calculations*: The molecules have been designed with the Avogadro package [18]. The optimization of the ground state geometry and the calculation of the electronic transitions have been performed with the package ORCA 3.0.3 [19], using the B3LYP functional [20] in the framework of the density functional theory. The Ahlrichs split valence basis set [21] and the all-electron nonrelativistic basis set SVPalls1 [22,23] have been employed. Moreover, the calculation utilizes the Libint library [24].

*Light transmission through multilayer photonic structures*: We employ the transfer matrix method to study the light transmission through the photonic structures [5]. The characteristic matrix of the $k$th layer, for a transverse electric (TE) light beam is given by

$$M_k = \begin{bmatrix} \cos\left(\frac{2\pi}{\lambda} n_k d_k\right) & -\frac{i}{n_k}\sin\left(\frac{2\pi}{\lambda} n_k d_k\right) \\ -i n_k \sin\left(\frac{2\pi}{\lambda} n_k d_k\right) & \cos\left(\frac{2\pi}{\lambda} n_k d_k\right) \end{bmatrix} \quad (1)$$

where $n_k$ is the refractive index and $d_k$ is the thickness of the layer. We underline that the refractive index can be a function of the wavelength. The matrix related to the whole multilayer structure is given by the product

$$M = M_1 \cdot M_2 \cdot \ldots \cdot M_k \cdot \ldots \cdot M_s = \begin{bmatrix} m_{11} & m_{12} \\ m_{21} & m_{22} \end{bmatrix} \quad (2)$$

where $s$ is the number of layers. From the $M$ matrix elements we can determine the transmission coefficient

$$t = \frac{2n_s}{(m_{11}+m_{12}n_0)n_s+(m_{21}+m_{22}n_0)} \quad (3)$$

In such expression $n_s$ relates to the refractive index of the substrate and $n_0$ relates to the refractive index of air. In this study, we choose $n_s = 1.46$, which corresponds to the refractive index of glass (neglecting its dependence upon wavelength). From the transmission coefficient we can derive the light transmission through the multilayer photonic crystal

$$T = \frac{n_0}{n_s}|t|^2 \tag{4}$$

**Results and Discussion**

In Figure 1 we show the light absorption spectrum, calculated with the density functional theory, of a photochromic molecule taken from Ref. [15] that in this work we call Irie1. Irie1a (solid black curve in Figure 1) refers to the open form of the molecule, while Irie1b (dashed-dotted red curve in Figure 1) refers to the close form of the molecule. The optimized geometries are reported in the Supporting Information.

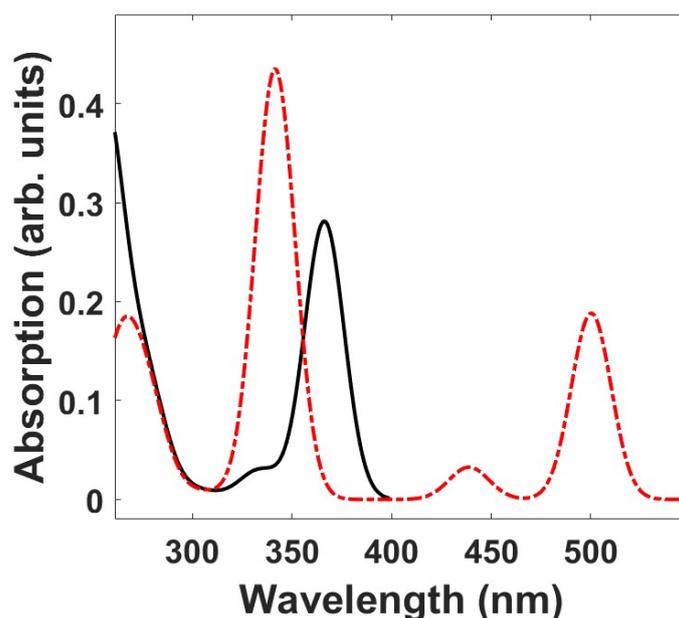

**Figure 1.** Calculated absorption of Irie1a (solid black curve) and Irie1b (dashed-dotted red curve).

In Figure 2 we show the calculated light absorption of a photochromic molecule taken from Ref. [16] that we call diarylethene1. Diarylethene1a (solid black curve in Figure 1) refers to the open form of the molecule, while diarylethene1b (dashed-dotted red curve in Figure 1) refers to the close form of the molecule. Also, for the diarylethene1 molecules we report in the optimized geometries in the Supporting Information. As expected, we observe the red shift of the lowest transition for the closed forms of Irie1 and diarylethene1.

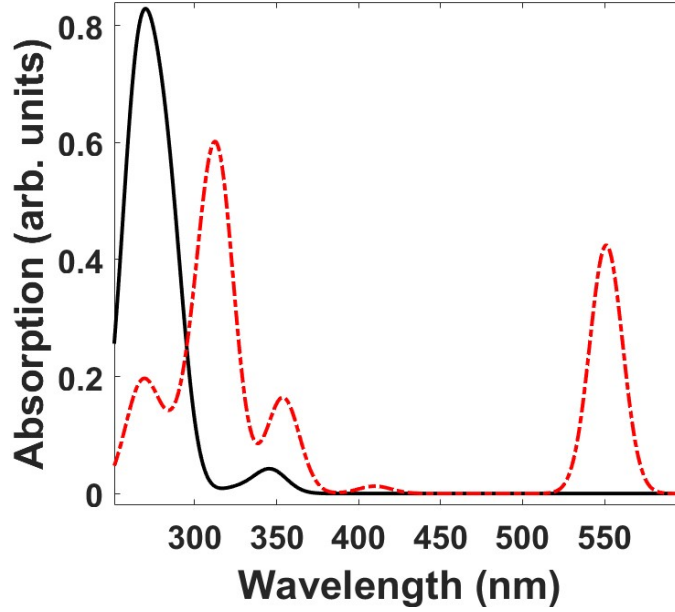

**Figure 2.** Calculated absorption of Diarylethene1a (solid black curve) and Diarylethene1b (dashed-dotted red curve).

From the absorption in Ref. [16] we can extract the oscillator strength [25]:

$$f_i = 4.319 \times 10^{-9} \int_{i-th\ band} \epsilon(\nu) d\nu \qquad (5)$$

with $\epsilon(\nu)$ the molar absorption of the *i*th band (fit from the experimental absorption curve). With $f_i$ we can extract the complex part of the dielectric permittivity [26]:

$$\varepsilon_2(\omega) = \frac{2\pi N_m e^2}{\varepsilon_0 m_e} \sum_i \frac{f_i}{\omega} D_i(\omega) \qquad (6)$$

where $e$ is the electron charge, $\varepsilon_0$ the vacuum dielectric permittivity, $m_e$ the electron mass, $N_m$ is the volume density of excited species, $D_i(\omega)$ the spectral shape of the fit bands from the absorption spectrum. Considering the refractive index and a complex function, we can recall the Kramers-Kronig relations [27]:

$$\eta(\tilde{\nu}) = n(\tilde{\nu}) + ik(\tilde{\nu}) \qquad (7)$$

$$n(\tilde{\nu}) - 1 = \frac{2}{\pi} \mathcal{P} \int_0^\infty \frac{\omega k(\omega)}{\omega^2 - \tilde{\nu}^2} d\omega \qquad (8)$$

$$k(\tilde{\nu}) - 1 = \frac{2\tilde{\nu}}{\pi} \mathcal{P} \int_0^\infty \frac{n(\omega) - 1}{\omega^2 - \tilde{\nu}^2} d\omega \qquad (9)$$

that can be written also for the dielectric permittivity:

$$\varepsilon(\omega) = \varepsilon_1(\omega) + \varepsilon_2(\omega) \qquad (10)$$

$$\varepsilon_1(\omega) = 1 + \frac{2}{\pi} \mathcal{P} \int_0^\infty \frac{\omega' \varepsilon_2(\omega')}{\omega'^2 - \omega^2} d\omega' \qquad (11)$$

$$\varepsilon_2(\omega) = -\frac{2\omega}{\pi} \mathcal{P} \int_0^\infty \frac{\varepsilon_1(\omega')}{\omega'^2 - \omega^2} d\omega' \qquad (12)$$

The Kramers-Kronig relationships are actually Hilbert transforms [28]. Thus, from the extracted imaginary part $\varepsilon_2(\omega)$ we can derivate the real part $\varepsilon_1(\omega)$. In the case of diarylethene, for real part of $\varepsilon$ we can used a constant offset of 2.25 as in Ref. [29].

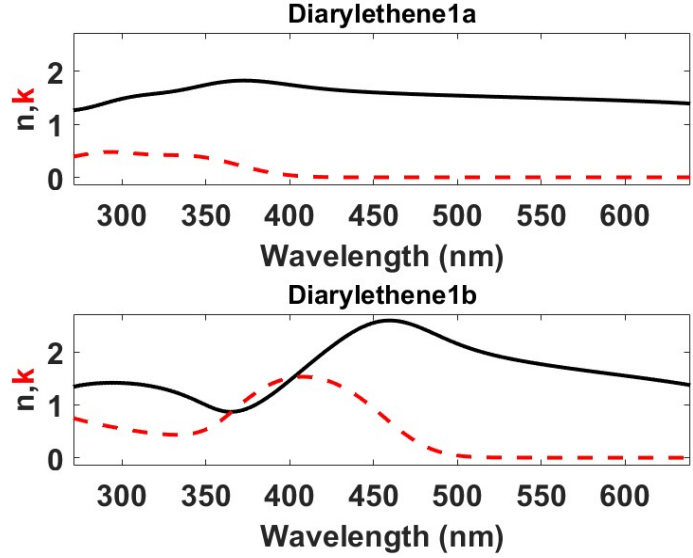

**Figure 3.** Real part of the refractive index function *n* (black solid curve) and imaginary part of the refractive index function *k* (red dashed curve) of diarylethene1a (open form, top) and of diarylethene1b (close form, bottom).

We design a one-dimensional multilayer photonic crystal by alternating layers of diarylethene1 with layers of poly vinyl carbazole (PVK).
The wavelength dependent refractive index of PVK is taken from Refs. [8,30,31] and can be written as

$$n^2 - 1 = \frac{0.09788^2}{\lambda^2 - 0.3257^2} + \frac{0.6901^2}{\lambda^2 - 0.1419^2} + \frac{0.8513\lambda^2}{\lambda^2 - 0.1417^2} \qquad (13)$$

with λ in micrometers.
The thickness of the diarylethene1 layers is 30 nm while the thickness of the PVK layers is 129 nm. The number of diarylethene1/PVK bilayers is 30. We show the light transmission of this multilayer photonic crystal in Figure 4, with the solid black curve corresponding to the crystal with diarylethene1a (open form) and the dotted-dashed red curve corresponding to the crystal with diarylethene1b (closed form). We observe the red shift of the photonic band gap.

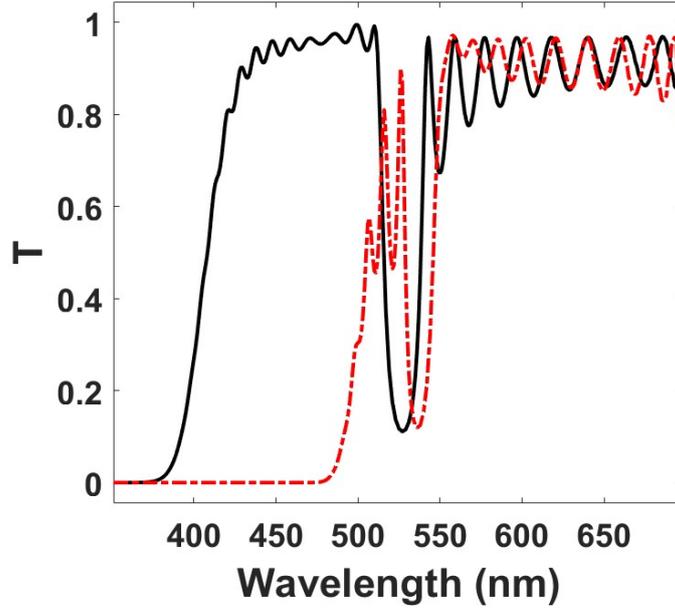

Figure 4. Light transmission through a multilayer photonic crystal made of 30 bilayers of Diarylethene1 and PVK. The solid black curve corresponds to the photonic crystal with diarylethene1a (open form), while the dotted-dashed red curve corresponds to the photonic crystal with diarylethene1b (closed form).

We also design a multilayer photonic crystal in which diarylethene1 layers are alternated with fluorine indium co-doped cadmium oxide (FICO) nanoparticle-based layers. We describe the plasmonic response of FICO nanoparticles with the Drude model [32]. Within the model, the frequency dependent complex dielectric permittivity of the material can be written as:

$$\varepsilon_{FICO,\omega} = \varepsilon_{1,\omega} + i\varepsilon_{2,\omega} \qquad (14)$$

In which

$$\varepsilon_{1,\omega} = \varepsilon_\infty - \frac{\omega_P^2}{(\omega^2 - \Gamma^2)} \qquad (15)$$

and

$$\varepsilon_{2,\omega} = \frac{\omega_P^2 \Gamma}{\omega(\omega^2 - \Gamma^2)} \qquad (16)$$

$\omega_P$ is the plasma frequency and $\Gamma$ is the free carrier damping. The plasma frequency is written as

$$\omega_P = \sqrt{\frac{Ne^2}{m^*\varepsilon_0}} \qquad (17)$$

In this expression $N$ is the number of charges, $e$ the electron charge, $m^*$ the effective mass and $\varepsilon_0$ the vacuum dielectric constant. For FICO, we use $N = 1.68 \times 10^{27} \ charges/m^3$ and $m^* = 0.43/m_0 \ kg$, a free carrier damping $\Gamma = 0.07$, and a high frequency dielectric constant, $\varepsilon_\infty = 5.6$, is taken from [33,34].

Since each layer of the photonic structure is made of necked nanoparticles, to determine the effective dielectric permittivity of the FICO:air layer (we call it $\varepsilon_{eff,\omega}$) we employ the Maxwell Garnett effective medium approximation [35,36]:

$$\varepsilon_{eff,\omega} = \varepsilon_{Air} \frac{2(1-f_{FICO})\varepsilon_{Air}+(1+2f_{FICO})\varepsilon_{FICO,\omega}}{2(2+f_{FICO})\varepsilon_{Air}+(1-f_{FICO})\varepsilon_{FICO,\omega}} \qquad (18)$$

in which $f_{FICO}$ is the filling factor of the FICO nanoparticles (in this study we choose a filling factor of 0.6). The photon energy dependent complex refractive index of the FICO:air layer is extracted considering that $n_{eff,\omega} = \sqrt{\varepsilon_{eff,\omega}}$.

In Figure 5 we show the light transmission of a multilayer photonic crystal of 50 bilayers of diarylethene1 and FICO nanoparticles. The transmission of the crystal with diarylethene1 in the open form corresponds to solid black curve, while the transmission of the crystal with diarylethene1 in the closed form corresponds to dotted-dashed red curve. Also with the diarylthene1/FICO crystal we observe a red shift. With respect to the diarylethene1/PVK crystal the transmission valleys change significantly between the open form and the closed form.

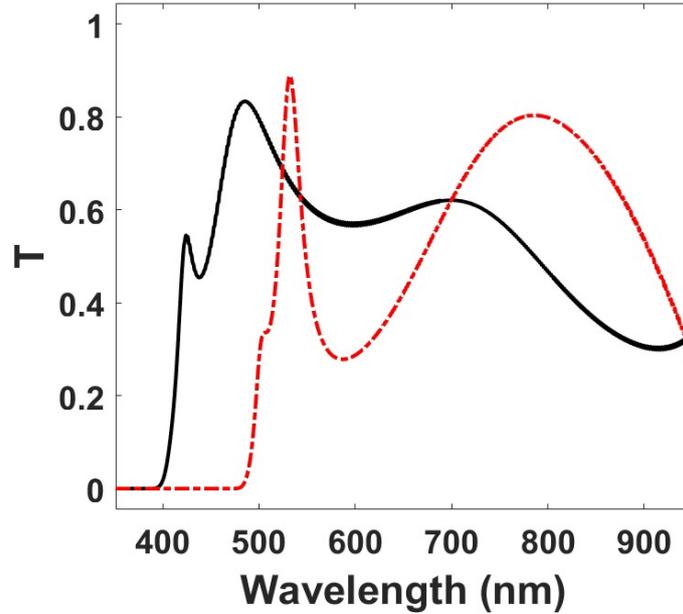

**Figure 5.** Light transmission through a multilayer photonic crystal made of 30 bilayers of Diarylethene1 and FICO nanoparticles. The solid black curve corresponds to the photonic crystal with diarylethene1a (open form), while the dotted-dashed red curve corresponds to the photonic crystal with diarylethene1b (closed form).

Multilayer photonic structures with an active material as layer can be very interesting for solid-state lasers. We calculate with DFT the absorption spectrum of BSB-Cz, one of the most promising organic materials for lasing [37,38,17]. In this calculation, vibrational levels are neglected. With respect to the experimental absorption spectrum of BSB-Cz as in Ref. [38], there is a mismatch with the electronic transitions at lower energies.

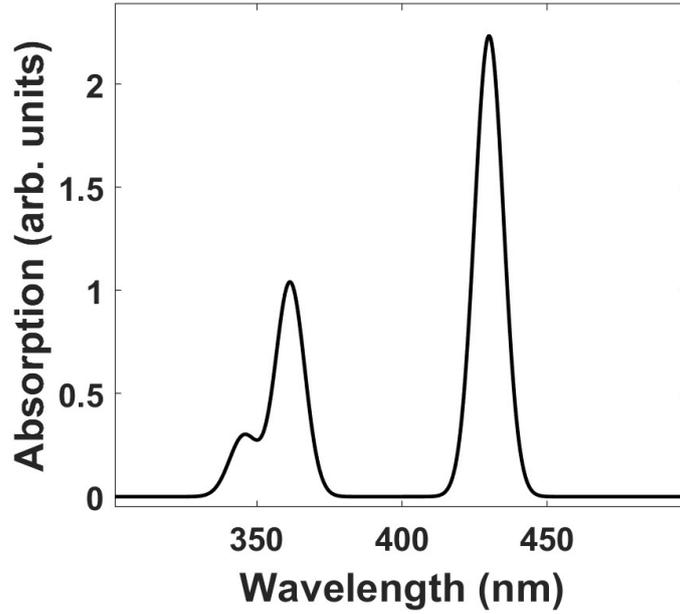
Figure 6. Calculated absorption spectrum of BSB-Cz.

We use the experimental complex refractive index of BSB-Cz measured in Ref. [38]. In this way we can design a 1D photonic crystal made by alternating layers of BSB-Cz and silicon dioxide. For silicon dioxide we use the following Sellmeier equation [39,40]

$$n^2 - 1 = \frac{0.6961663\lambda^2}{\lambda^2 - 0.0684043^2} + \frac{0.4079426\lambda^2}{\lambda^2 - 0.1162414^2} + \frac{0.8974794\lambda^2}{\lambda^2 - 9.896161^2} \tag{19}$$

The photonic crystal is made of 10 bilayers of BSB-Cz and $SiO_2$. The thickness of the BSB-Cz layers is 11 nm while the thickness of $SiO_2$ layers is 260 nm.

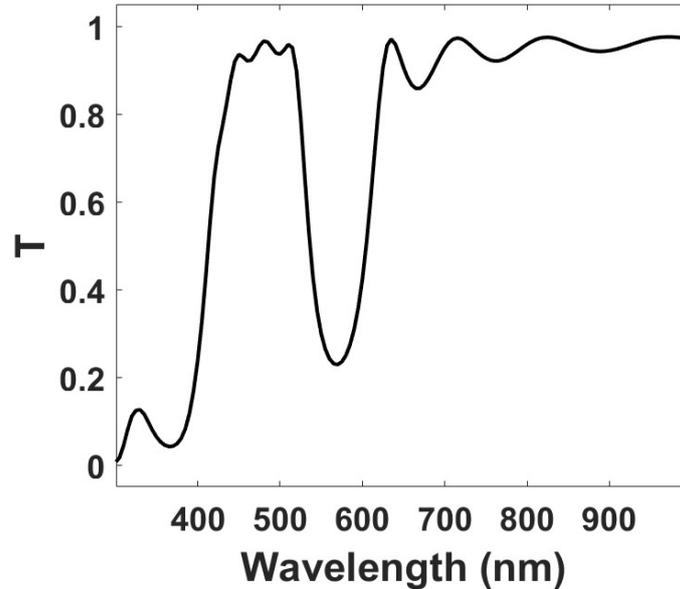
Figure 7. Calculated transmission spectrum of BSB-Cz/$SiO_2$ multilayer.

We design also a microcavity in which a BSB-Cz layer is placed between two photonic crystals made of 30 bilayers of ITO and FICO nanoparticle-based layers [(ITO/FICO)30/BSB-Cz/(ITO/FICO)30].

We describe also the plasmonic response of ITO nanoparticles with the aforementioned Drude model. For ITO, we use $N = 2.49 \times 10^{26}\ charges/m^3$ and $m^* = 0.4/m_0\ kg$, and a free carrier damping $\Gamma = 0.1132$ [34,41]. The value of the high frequency dielectric constant, $\varepsilon_\infty = 4$, is taken from [35]. To determine the effective dielectric permittivity of the ITO:air layer we employ the Maxwell Garnett effective medium approximation as for FICO:air layers. In this structure, for both ITO:air layers and FICO:air layers we use a filling factor of 0.65.

The thickness of the BSB-Cz film is 62.2 nm, the ITO layer thickness is 88.74 nm and the FICO layer thickness is 62.61 nm. We design the microcavity in order to have the cavity mode at 480 nm, where lasing of BSB-Cz has been demonstrated (with indications of current-injection lasing) [17]. We choose ITO and FICO nanoparticle-based layers since they have a good conductivity [42,43] and they could in principle ensure charge injection in BSB-Cz.

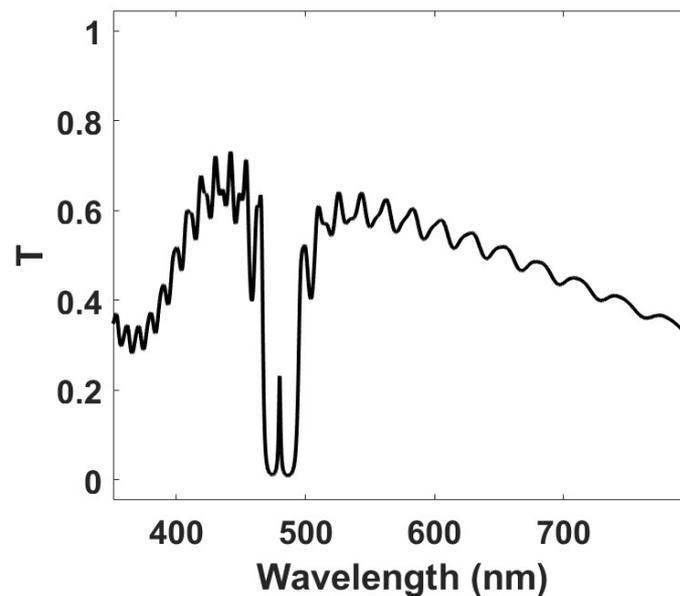

Figure 8. Light transmission spectrum of a (ITO/FICO)$_{30}$/BSB-Cz/(ITO/FICO)$_{30}$.

**Conclusion**

In this work we study the light transmission properties of one-dimensional multilayer photonic structures that include layers of molecules with interesting properties as photochromism and optical gain. After studying via DFT the absorption spectra of two photochromic molecules (Irie1 and diarylethene1), we have designed multilayer photonic crystals alternating layers of diarylethene1 and PVK, or diarylethene1 and FICO. In these structures we observe a shift of the photonic features for the open form and close form of diarylethene1. We have also designed a multilayer photonic crystal alternating BSB-Cz and SiO$_2$, and a microcavity in which BSB-Cz is sandwiched between two photonic crystals made with ITO and FICO nanoparticle-based layers. These photonic structures can be interesting for colour switches, in the case of the photochromic molecules, and for solid-state lasers, in the case of BSB-Cz.

**Optimized geometry of BSB-Cz**

| | | | |
|---|---|---|---|
| C | -5.18707731845260 | 0.70580425173054 | 1.40687734354036 |
| C | -5.95943591341883 | 0.94232696292358 | 2.55364858877459 |
| C | -5.32568618797214 | 1.34126865621023 | 3.73047615321365 |
| C | -3.92659074480217 | 1.49656042396041 | 3.77160185919136 |
| C | -3.13522706537972 | 1.26620365416615 | 2.64459404549157 |
| C | -3.77749660147309 | 0.87935976951844 | 1.46091845517512 |
| H | -5.91689448235239 | 1.53112920477760 | 4.63132605774338 |
| H | -7.04559644894551 | 0.80971288217600 | 2.52563830174757 |
| H | -2.04980623711333 | 1.38046636300795 | 2.69070072233018 |
| H | -3.44608305620983 | 1.80214057785020 | 4.70638120740647 |
| N | -3.23094266668557 | 0.56934150027348 | 0.21041159930255 |
| C | -5.50393364068408 | 0.27251786459674 | 0.05996528491464 |
| C | -4.27309519864638 | 0.19848797135400 | -0.64697908380119 |
| C | -6.70185254161544 | -0.04134602906550 | -0.59900997797995 |
| C | -6.66157551579837 | -0.41661406250093 | -1.94173423050408 |
| C | -5.43555044248494 | -0.47175829115210 | -2.63252388001231 |
| C | -4.22955153475036 | -0.16366737138027 | -1.99994116027805 |
| H | -3.28527893312307 | -0.19965492050381 | -2.54750383442121 |
| H | -5.42457793931794 | -0.76076668643500 | -3.68794385938739 |
| H | -7.58878718359770 | -0.66685279160020 | -2.46552358923188 |
| H | -7.65597286755428 | 0.01069638017668 | -0.06569329991248 |
| C | -1.85352995002194 | 0.60422173845912 | -0.12402878766362 |
| C | -1.10288885028165 | 1.77923452738335 | 0.05895376417077 |
| C | 0.25127352809385 | 1.81164391260046 | -0.25984520557878 |
| C | 0.90983304714368 | 0.67983599789871 | -0.78881793019137 |
| C | -1.21529080015232 | -0.53158573951601 | -0.64396630996118 |
| C | 0.13757364158427 | -0.48584344369390 | -0.97626058881045 |
| H | -1.59532778434736 | 2.67606872245329 | 0.44368375691409 |
| H | 0.80169302924926 | 2.74344792724224 | -0.10679059799973 |
| H | 0.61697156399555 | -1.38341159047333 | -1.38018789362604 |
| H | -1.78347513141342 | -1.45578337327029 | -0.77692972266258 |
| C | 2.33121687172815 | 0.65563061569142 | -1.14657172346964 |
| C | 3.22677545331294 | 1.66204661990228 | -1.01307013189724 |
| C | 4.64741475399804 | 1.63673091529344 | -1.37069406495583 |
| H | 2.88692487372000 | 2.61603316153213 | -0.59315625669948 |
| H | 2.67252339833523 | -0.29916157194878 | -1.56282703933980 |
| C | 5.42874017098554 | 2.79210415136675 | -1.16081904355449 |
| C | 6.78381235534812 | 2.82755839304278 | -1.48190369725215 |
| C | 7.43398302508963 | 1.70448619854090 | -2.02850014016059 |
| C | 5.30063812344980 | 0.51259093123478 | -1.92153767953623 |
| C | 6.65408391458098 | 0.54663236653147 | -2.23835319536072 |
| H | 7.35501835127445 | 3.73728787011160 | -1.27684533049873 |
| H | 4.95921477313451 | 3.68091003525062 | -0.72627375420780 |
| H | 7.11420533439316 | -0.33889131082096 | -2.68543221654799 |
| H | 4.74006123657194 | -0.40558328438989 | -2.11562215239871 |
| C | 8.87698101316561 | 1.73518041765680 | -2.36793250585661 |
| C | 9.49493198238240 | 2.91301375713158 | -2.84121337028915 |
| C | 10.84784448175634 | 2.94835289966773 | -3.16039279000965 |
| C | 11.66464393203350 | 1.80506676543754 | -3.01993855473759 |
| C | 9.68892156937881 | 0.59285517647783 | -2.23142400812266 |
| C | 11.04521335193462 | 0.62940946195885 | -2.54742538107408 |
| H | 11.27410065029529 | 3.88265457895735 | -3.53560197178199 |
| H | 8.89300781649238 | 3.81406154488741 | -2.98918342089497 |
| H | 11.64925078523340 | -0.27473088883343 | -2.41665777917596 |
| H | 9.25839227541425 | -0.33449653345883 | -1.84418167762564 |
| C | 13.09526830885627 | 1.77952965581637 | -3.33425244999279 |
| C | 13.85436579946473 | 2.80741011797400 | -3.78140105572590 |
| C | 15.28541946445925 | 2.77879363708028 | -4.09725060895656 |
| H | 13.38028501619674 | 3.78426866780981 | -3.93212114577696 |
| H | 13.57218215696070 | 0.80531448487329 | -3.17484836554766 |
| C | 15.92084230236175 | 3.97010974624433 | -4.50564715372004 |

```
C   16.08348739477332      1.61615819811546     -4.01895018520352
C   17.43786543690588      1.64411145182162     -4.33623584711909
C   18.05656176990686      2.84389569587276     -4.73106627129856
C   17.28113865688753      4.01040115932469     -4.80833167867815
H   17.75219888665529      4.95337198401675     -5.09650940252150
H   15.33278230882613      4.89133138688389     -4.57261367098203
H   18.02886588730133      0.72562987741889     -4.29232853466495
H   15.63625436805351      0.66460402161833     -3.72012349300733
N   19.43817487324469      2.87069886133915     -5.04778028027400
C   19.99486156981142      3.32647912340135     -6.24863707883224
C   20.47695487413745      2.43190865042106     -4.21869224263225
C   21.71532245759846      2.60827333668779     -4.89366700301419
C   21.40723808059160      3.17869040120981     -6.19054555790902
C   19.36086887543387      3.82604525816305     -7.39402025883937
C   22.19022347525789      3.55677548960683     -7.29123563970478
C   20.16257324362054      4.19529034980413     -8.47586711826850
C   21.56427420009666      4.06796726943046     -8.42807263601668
H   19.68833730472841      4.58959607969117     -9.38023000168112
H   18.27380186345902      3.92068788518442     -7.44602278809049
H   23.27886949140858      3.44689877424785     -7.25915478121197
H   22.16369802284978      4.36841260804637     -9.29265797232322
C   20.42379613689798      1.92564009516715     -2.91305438468084
C   21.62789785942206      1.57721374696014     -2.29818408203929
C   22.86149304750244      1.73209374724269     -2.95976446435496
C   22.91116505196736      2.25065963877682     -4.25359633499374
H   23.78703009985977      1.44725963767349     -2.45103582120001
H   23.87092412515419      2.38082916446774     -4.76277988358321
H   19.47391802473070      1.81171200830286     -2.38640587765987
H   21.60954929713639      1.17840052891592     -1.27944523194156
```

**Optimized Geometry of Irie-1a**

| | | | |
|---|---|---|---|
| C | -3.474210 | 1.728310 | 0.316400 |
| C | -2.207260 | 1.891170 | -0.160970 |
| C | -3.979540 | 3.108100 | 0.742240 |
| C | -1.921190 | 3.385270 | -0.220280 |
| C | -3.297730 | 3.996280 | -0.273920 |
| F | -1.159430 | 3.784800 | -1.265780 |
| F | -1.310090 | 3.816110 | 0.917770 |
| F | -3.554000 | 3.380960 | 2.006590 |
| F | -5.310060 | 3.336100 | 0.748900 |
| F | -3.859360 | 3.854420 | -1.507490 |
| F | -3.327830 | 5.310510 | 0.025650 |
| C | -1.146630 | 1.007980 | -0.605080 |
| C | -4.330370 | 0.599020 | 0.597400 |
| C | 0.303250 | 1.234650 | -0.370430 |
| C | 1.085490 | 0.222170 | -0.936630 |
| C | -1.340450 | -0.129770 | -1.379990 |
| S | 0.120130 | -0.979400 | -1.668070 |
| C | -2.557500 | -0.731220 | -2.006550 |
| H | -2.419520 | -0.807610 | -3.091530 |
| H | -3.456500 | -0.138520 | -1.847660 |
| H | -2.732260 | -1.741150 | -1.617640 |
| C | 1.038750 | 2.267200 | 0.286820 |
| C | 2.500330 | 0.231240 | -0.889430 |
| C | 3.166190 | 1.295130 | -0.286670 |
| C | 2.439270 | 2.308260 | 0.301310 |
| H | 3.071580 | -0.575560 | -1.341440 |
| H | 4.252890 | 1.326930 | -0.276790 |
| H | 0.550130 | 3.091180 | 0.786820 |
| H | 2.948310 | 3.145670 | 0.774250 |
| C | -3.910500 | -0.565720 | 1.226970 |
| C | -2.565280 | -1.016950 | 1.697280 |
| H | -2.126410 | -1.709220 | 0.970980 |
| H | -2.639220 | -1.547580 | 2.653230 |
| H | -1.881860 | -0.181320 | 1.854250 |
| C | -5.781430 | 0.554700 | 0.324760 |
| C | -6.378460 | -0.570370 | 0.901090 |
| S | -5.206250 | -1.646870 | 1.520250 |
| C | -6.667250 | 1.446880 | -0.350390 |
| C | -8.059450 | 1.295650 | -0.284640 |
| C | -8.611020 | 0.210680 | 0.370710 |
| C | -7.779170 | -0.755490 | 0.931980 |
| H | -8.707030 | 2.044380 | -0.736150 |
| H | -6.284660 | 2.307890 | -0.892820 |
| H | -9.691300 | 0.101780 | 0.435510 |
| H | -8.209380 | -1.630490 | 1.412500 |

**Optimized Geometry of Irie-1b**

| | | | |
|---|---|---|---|
| C | -3.60266 | 2.07204 | 0.13490 |
| C | -2.13721 | 2.25378 | -0.00684 |
| C | -4.23161 | 3.44433 | 0.12280 |
| C | -1.90469 | 3.74246 | 0.12804 |
| C | -3.16513 | 4.33437 | -0.47810 |
| F | -0.81929 | 4.22240 | -0.52213 |
| F | -1.76997 | 4.08202 | 1.43493 |
| F | -4.51868 | 3.82533 | 1.39638 |
| F | -5.38704 | 3.52500 | -0.57989 |
| F | -3.14347 | 4.20423 | -1.83786 |
| F | -3.33722 | 5.64762 | -0.21729 |
| C | -1.33230 | 1.20099 | -0.27913 |
| C | -4.14746 | 0.84972 | 0.31112 |
| C | 0.14683 | 1.09649 | -0.29738 |
| C | 0.60263 | -0.22598 | -0.40484 |
| C | -1.99930 | -0.16841 | -0.61210 |
| S | -0.67854 | -1.44090 | -0.44609 |
| C | -2.36391 | -0.15487 | -2.12150 |
| H | -1.51586 | 0.15679 | -2.74535 |
| H | -3.17848 | 0.54568 | -2.33804 |
| H | -2.68162 | -1.13944 | -2.48327 |
| C | 1.10421 | 2.10077 | -0.17343 |
| C | 1.95689 | -0.55025 | -0.43441 |
| C | 2.89733 | 0.46962 | -0.34776 |
| C | 2.47127 | 1.78976 | -0.21109 |
| H | 2.27870 | -1.58437 | -0.51874 |
| H | 3.95893 | 0.23758 | -0.37241 |
| H | 0.84680 | 3.14051 | -0.03400 |
| H | 3.20650 | 2.58746 | -0.12552 |
| C | -3.23758 | -0.40284 | 0.32914 |
| C | -2.84689 | -0.68494 | 1.80195 |
| H | -2.33861 | -1.64830 | 1.92044 |
| H | -3.72098 | -0.70202 | 2.46576 |
| H | -2.17655 | 0.08791 | 2.19580 |
| C | -5.57325 | 0.48091 | 0.42030 |
| C | -5.80622 | -0.87287 | 0.14360 |
| S | -4.33661 | -1.79622 | -0.20015 |
| C | -6.65310 | 1.28527 | 0.77660 |
| C | -7.94534 | 0.74669 | 0.80131 |
| C | -8.16201 | -0.59729 | 0.48800 |
| C | -7.08503 | -1.41823 | 0.16320 |
| H | -8.78802 | 1.38047 | 1.06967 |
| H | -6.52928 | 2.32678 | 1.05204 |
| H | -9.16967 | -1.00509 | 0.50920 |
| H | -7.24519 | -2.46842 | -0.06252 |

**Optimized Geometry of Diarylethene 1a**

| | | | |
|---|---|---|---|
| C | -1.87470541770156 | 1.59165856138569 | -0.03615518763699 |
| C | -3.79586309087645 | 2.90463845245704 | 0.36979422437935 |
| C | -1.45972494664256 | 3.04082771650044 | -0.22893673064555 |
| C | -2.80359147616385 | 3.78767064026008 | -0.41850225481685 |
| F | -0.63753513641960 | 3.20784813249107 | -1.28546821017008 |
| F | -0.82981393830126 | 3.56439356774384 | 0.85955710875699 |
| F | -3.84092233124772 | 3.31191329541820 | 1.66806073259206 |
| F | -5.05200823198005 | 3.01208083439127 | -0.11633947025046 |
| F | -3.14213536208815 | 3.77268915556931 | -1.72799685006469 |
| F | -2.76755517920284 | 5.05854582804431 | 0.00297825640432 |
| C | -0.92412955900461 | 0.47988785607397 | -0.26207399353880 |
| C | -3.99798249624876 | 0.31953837052663 | 0.60435313415529 |
| C | 0.28257582368626 | 0.24207847293770 | 0.50884365010969 |
| C | 0.96439543048151 | -0.87859730772095 | 0.07261119241938 |
| C | -1.10762696940493 | -0.43412843655091 | -1.28647010659036 |
| S | 0.16296213349000 | -1.60676641687945 | -1.29875005115416 |
| C | -2.19516201543756 | -0.50268985261681 | -2.31757698122279 |
| H | -1.83005954834166 | -0.96186937682926 | -3.25175552397836 |
| H | -2.56370001423871 | 0.50746430373018 | -2.56100364255124 |
| H | -3.05690840275563 | -1.09899404312896 | -1.96526373419956 |
| C | 0.77626263161145 | 1.15103497010002 | 1.60510628613115 |
| C | -3.72643844554084 | -0.45124370861879 | 1.72278596845075 |
| C | -2.65905078453872 | -0.26835937251397 | 2.76057663011016 |
| H | -1.73734047408616 | -0.82040383524552 | 2.49886912234660 |
| H | -2.99450743346213 | -0.62687760845042 | 3.74892016468531 |
| H | -2.39652449856184 | 0.79772601183013 | 2.85816712183779 |
| C | -5.13659968503547 | -0.14907010730884 | -0.15864218368365 |
| C | -5.69737544749212 | -1.28845773968493 | 0.39340633409054 |
| S | -4.85070331557889 | -1.75837477258022 | 1.84935497193798 |
| C | -5.61040443777445 | 0.48694903985398 | -1.44251022633913 |
| C | -8.00542052080101 | -1.50687481126849 | -0.58032580711387 |
| C | -6.83456153125038 | -2.10521012821300 | -0.07191852560038 |
| H | 1.10298123458767 | 2.12172706288822 | 1.19522714412370 |
| H | 1.63495007124559 | 0.71047658543520 | 2.13256732524873 |
| H | -0.01047783943812 | 1.37128260140537 | 2.34733970002006 |
| C | 2.20726169994351 | -1.50094031335371 | 0.58055527389740 |
| C | 2.34299909406850 | -1.85300175655437 | 1.93796451943855 |
| C | 3.27478252483805 | -1.79030648100069 | -0.29201664065254 |
| C | 4.44041515966031 | -2.40083736154168 | 0.17793669855899 |
| C | 4.56597775387525 | -2.73230361025020 | 1.53090980023302 |
| C | 3.51228927403276 | -2.45607155418135 | 2.40869791725112 |
| H | 3.19020742539256 | -1.52034252143070 | -1.34884413793171 |
| H | 5.47960765230499 | -3.20885626427117 | 1.89947266477854 |
| H | 5.25816530703475 | -2.61424702446853 | -0.51770190634821 |
| H | 3.59460727571208 | -2.72502868737424 | 3.46650593865241 |
| H | 1.51227081120288 | -1.66932007644312 | 2.62495087019007 |
| H | -4.80158607790448 | 1.05329021360519 | -1.93067006868330 |
| H | -6.43324638414748 | 1.20149010390036 | -1.26889528412786 |
| H | -5.97443129694977 | -0.27585137421994 | -2.15114325958189 |
| C | -9.08001858644432 | -2.28707204158444 | -1.01431463900846 |
| H | -8.08541164032925 | -0.41776560365937 | -0.61225532562239 |
| C | -9.01593197507580 | -3.68220373238310 | -0.94217859963320 |
| C | -6.78773903667803 | -3.51345819777345 | -0.00220851172673 |
| C | -7.86541438085987 | -4.29137222246836 | -0.42948059347647 |
| H | -9.85974236390264 | -4.29140128148201 | -1.28067707158783 |
| H | -9.97988829825281 | -1.79861232583535 | -1.40163493167235 |
| H | -7.80234489455192 | -5.38270373734815 | -0.36762575580293 |
| H | -5.88618925890722 | -4.00127782605061 | 0.38043463141071 |

**Optimized Geometry of Diarylethene 1b**

| | | | |
|---|---|---|---|
| C | -3.21055030257768 | 1.54785288836691 | -0.33092702267615 |
| C | -1.76265532443919 | 1.65743736941559 | -0.11750640018324 |
| C | -3.77248631909801 | 2.90176649931283 | -0.68443730948350 |
| C | -1.37677208213706 | 3.11089133051334 | -0.11027144841351 |
| C | -2.52801791442704 | 3.81467941873579 | -0.87852536217940 |
| F | -0.19765201359465 | 3.39217761911785 | -0.71542595693655 |
| F | -1.30780446703447 | 3.63229006869620 | 1.15244819981019 |
| F | -4.55647027257752 | 3.41966559455034 | 0.30205571575178 |
| F | -4.51413306105689 | 2.92318492818453 | -1.82482194661368 |
| F | -2.20884350178183 | 3.85060233638210 | -2.18952131219957 |
| F | -2.74137416693649 | 5.06865590343090 | -0.45499250638770 |
| C | -0.96621946561427 | 0.54911955487253 | 0.04273870649341 |
| C | -3.88361100328760 | 0.36736774533282 | -0.13312924275011 |
| C | 0.40587906641223 | 0.39397652000795 | 0.47758205285064 |
| C | 0.76686693044308 | -0.92672373307524 | 0.65174881689609 |
| C | -1.65122243781105 | -0.78532940342468 | -0.29173557755194 |
| S | -0.51092356363555 | -2.10680200805814 | 0.34756311661632 |
| C | -1.70574809096903 | -0.95063171131418 | -1.83040113324503 |
| H | -0.68383451706759 | -0.90637453557175 | -2.23997503304585 |
| H | -2.29864263545910 | -0.14267344450484 | -2.29023005658613 |
| H | -2.15594891649885 | -1.91625750712931 | -2.10932988652269 |
| C | 1.29853629337977 | 1.54519512891310 | 0.86944619933727 |
| C | -3.03400989875205 | -0.80081276387617 | 0.39034648257660 |
| C | -2.96449547331582 | -0.68532060573088 | 1.93584081415137 |
| H | -2.39972091234392 | -1.52408894090553 | 2.37037606906527 |
| H | -3.98529885683076 | -0.69461532304265 | 2.35025797002186 |
| H | -2.48036810612544 | 0.25999720559412 | 2.23395616914751 |
| C | -5.28400190880441 | 0.00503066315402 | -0.26783300960326 |
| C | -5.48803344188267 | -1.35580534732725 | -0.19887618228751 |
| S | -4.02494628127763 | -2.32935256509915 | 0.00915303733230 |
| C | -6.39058555431167 | 0.99762225884583 | -0.51486677485201 |
| C | -7.86737934902837 | -1.75415172867127 | 0.52395896325400 |
| C | -6.76180106433049 | -2.10205520712233 | -0.27760091519209 |
| H | 1.72037911348860 | 2.06720125613949 | -0.00576830239147 |
| H | 2.13183709898187 | 1.20155415379776 | 1.50070377676196 |
| H | 0.72920690067797 | 2.29523277679742 | 1.44216473093493 |
| C | 2.07373493414834 | -1.46424215448779 | 1.07821817226587 |
| C | 2.14530705677315 | -2.51288656055723 | 2.01847015829153 |
| C | 3.27695698176526 | -0.97481073918996 | 0.53044907415296 |
| C | 4.50783317867532 | -1.50467530123745 | 0.92231585862658 |
| C | 4.56349391886214 | -2.53424026767105 | 1.86745155671187 |
| C | 3.37721629752674 | -3.03701632504972 | 2.41232969749374 |
| H | 3.24345552188038 | -0.18825226334855 | -0.22762190096527 |
| H | 5.52870378187672 | -2.94796216676768 | 2.17435648854013 |
| H | 5.42941181199901 | -1.11511875073255 | 0.47962565857025 |
| H | 3.41099865161802 | -3.84427067512756 | 3.15072725785564 |
| H | 1.22358224366045 | -2.90827603919477 | 2.45484341234884 |
| H | -6.19154168062594 | 1.58826250273371 | -1.42162065370400 |
| H | -6.48003169580482 | 1.71774088358595 | 0.31775989072470 |
| H | -7.35971721094936 | 0.49628073854568 | -0.64734332211549 |
| C | -9.05606978173788 | -2.48462681805308 | 0.45594411645511 |
| H | -7.78539203468640 | -0.91724838151600 | 1.22221561710087 |
| C | -9.16334660825213 | -3.57843540063517 | -0.40938827933120 |
| C | -6.87858445819063 | -3.21674846141640 | -1.13233207940761 |
| C | -8.06909804114617 | -3.94325828989920 | -1.20098755365438 |
| H | -10.09588816802457 | -4.14882553543814 | -0.46197866617512 |
| H | -9.90227294856714 | -2.20069447715661 | 1.08936177490425 |
| H | -8.14207850851407 | -4.80051532490903 | -1.87744345897481 |
| H | -6.02871774266283 | -3.50497658778544 | -1.75802826161455 |